# Demonstration of 3D ISAR Security Imaging at 24GHz with a Sparse MIMO Array

Zhanyu Zhu and Feng Xu*, *Senior Member, IEEE*

*Abstract*- **A 3D ISAR security imaging experiment at 24GHz is demonstrated with a sparse MIMO array. The MIMO array is an 8Tx/16Rx linear array to achieve real-aperture imaging along the vertical dimension. It is time-switching multiplexed with a low-cost FMCW transceiver working at 22GHz-26GHz. A calibration procedure is proposed to calibrate the channel imbalance across the MIMO array. The experiment is conducted on human moving on a cart, where we take advantage of the linear motion of human to form inverse synthetic aperture along the horizontal dimension. To track the motion of human, a 3D depth camera is used as an auxiliary sensor to capture the rough position of target to aid ISAR imaging. The back projection imaging algorithm is implemented on GPU for quasi-real-time operation. Finally, experiments are conducted with real human with concealed objects and a preliminary automatic object recognition algorithm based on convolutional neural networks are developed and evaluated on real data.**

## I. INTRODUCTION

Microwave and millimeter wave security imaging of human body has important applications in monitoring of security threats and detection of contraband smuggling. The former is a major issue of anti-terrorist activities in public areas such as airports, schools, governments and public events. The latter is a common problem for customs and border protection. With the increasingly terrorism around the world and the frequent occurrence of terrorist attacks in public places, anti-terrorism technology has become one of the major demands, and there is an urgent need to develop a new type of security screening technology for human. The screening technology should have desired features like small size, light weight and deployable in a temporal site. Ideally, it should also be able to perform efficient screening without sacrificing the throughput of human traffic.

The most successful security screen technology is the airport security check before boarding. Metal detector [1], millimeter-wave holographic imaging [2][3] and X-ray imaging [4][5] technology are widely used in this scenario. Metal detectors do not perform imaging and only detect the presence of metal objects, which can only be used as a supplemental measure. X-ray technology is widely used for luggage screening but not for human body due to its nature of ionizing radiation. Microwave or millimeter-wave holographic imaging technology [6] is one of the promising technologies for human body security screening, because these working frequencies are non-ionizing radiation. Besides, the products [6] also come with automatic target recognition capability which removes the privacy concerns. However, the holographic imaging technology is not efficient enough for many other scenarios. It employs mechanical scanning which requires the person under screening to stand still.

Multi-input multi-output (MIMO) sparse array is viable technology to replace mechanical scanning. To address this need, a fast imaging system based on multi-input multi-output (MIMO) 2D sparse array [7][8][9] was developed. The system integrates 3072 TX channels and 3072 RX channels, in which the TX channels transmit signals sequentially in time. Although the number of array elements has been greatly reduced by using sparse array, there are still thousands of receiving channels which makes the system extremely complex. In order to further reduce the number of antennas or channels, we have to replace the 2D real aperture imaging mechanism with a novel architecture.

In this paper, we propose to further reduce the system complexity and size-weight by using inverse synthetic aperture imaging in the horizontal dimension. It takes advantage of linear motion of human body on a transportation platform to form a synthetic aperture. The radar system can be reduced to a 1D sparse MIMO array for real aperture imaging only in the vertical dimension. Given that radar does ranging along the 3rd dimension of depth, a 3D inverse synthetic aperture radar (ISAR) imaging ability can be achieved. The radar system works at 24GHz with a bandwidth of 4GHz. It is equipped with a 1D sparse array of 8Tx and 16Rx antennas. The 8Tx antennas are time-division multiplexed to 1 transmit channel via a 1-to-8 switch, while the 16Rx antennas are time-division multiplexed to 1 receive channel via a 16-to-1 switch. Therefore, the radar transceiver is a low-cost single channel transceiver employing the simple frequency modulated continuous wave (FMCW) architecture. A novel array calibration approach is developed to calibrate the channel imbalance across the MIMO array. To tackle the challenge of ISAR focusing, auxiliary human motion tracking technology based on depth camera is used to obtain the real-time human body motion parameters, which are then used to focus the ISAR image. Moreover, a quasi-real-time 3D backprojection (BP) algorithm is implemented with acceleration of graphic processing unit (GPU). Finally, experiments are carried out on real person with concealed

Manuscript received xxxx

This work was supported in part by National Key Research & Development Program of China no. 2017YFB0502703, Natural Science Foundation of China no. 61822107, 61571134 and the SAST Research Fund no. SAST2017-078.

Zhanyu Zhu and Feng Xu are with the Key Laboratory for Information Science of Electromagnetic Waves (MoE), Fudan University, Shanghai 200433, China (* corresponding to fengxu@fudan.edu.cn)

objects. A preliminary object recognition algorithm based on deep neural networks are developed and tested.

The remainder of the paper is organized as follows. Section 2 describes the radar system design with system architecture, antenna design, baseband subsystem, RF subsystem, and system timing. Section 3 illustrates the imaging algorithm which includes the array calibration, 3D backpropagation on GPU and auxiliary sensor for motion tracking. Section 4 demonstrates the experiment results of human imaging and object recognition. Finally, this paper is concluded in section 5.

## II. RADAR SYSTEM

### A. System Architecture

We set up a system for 3D ISAR security imaging at 24GHz with a linear sparse MIMO array, whose architecture is depicted in Fig. 1. The radar system interfaces with a computer which runs the user interface (UI) control and the signal and image processing algorithms. The computer is connected with an auxiliary 3D depth camera which is used to capture human motion information. The radar system is a low-cost implementation where a FMCW transceiver is interfaced with a sparse MIMO array. The radar transmit channel consists of a chirp waveform generator, an 8X multiplier chain, a radio frequency (RF) power amplifier and a 1-to-8 RF switch which is time-multiplexed with the 8 Tx antennas. The radar receiver channel consists of 16 Rx antennas which are time-multiplexed via a 16-to-1 RF switch to the direct downconverter mixer, and then digitized by an analogy-to-digital converter (ADC) before sent into the FPGA-based control and data buffering system. The received data are finally sent to the computer for ISAR processing and object recognition. Note that the computer sends commands to the FPGA-based control system who coordinates the radar system working in real time. Both the radar raw data and the camera data are collected by the computer where they are fused to form focused ISAR image.

The system parameters are listed in TABLE I. The FMCW mechanism is adopted in this system.

TABLE I
The system operating parameters

| parameter | value | parameter | value |
|---|---|---|---|
| Waveform | FMCW | Frequency | 24GHz |
| Bandwidth | 4GHz | Range resolution | ~3.75cm |
| Equivalent array size | ~100cm * 70cm | Antenna Beam width | ~50deg * 50deg |
| Horizontal resolution | ~1.2cm | Vertical resolution | ~1.8cm |
| Tx array | 8(unit) | Rx array | 16(unit) |
| Tx power | 13dBm | Pulse width | 30us |

### B. Antenna Design

The antenna array is composed of 16 receiving antennas in the middle segment and 8 transmitting antennas on the upper and lower ends, i.e. 4 on each end. So, an equivalent 128 virtual channels are formed through MIMO technology. The spacing of Tx and Rx antennas are designed in such a way that the 128 virtual elements are evenly distributed across the entire aperture of 1m length. In order to align the field of view (FoV) of all antennas, the array manifold is designed as an arc whose origin collocates with the scene center which is in 1.5m away from the array. The manifold design diagram is shown in Fig. 2. Given the radiation pattern of the adopted 10dB standard-gain horn antenna, the FoV coverage of the entire array can be inferred, as shown in Fig. 2. The center of the FoV is 1.5m away and it covers about 1x1m in vertical and horizontal dimensions.

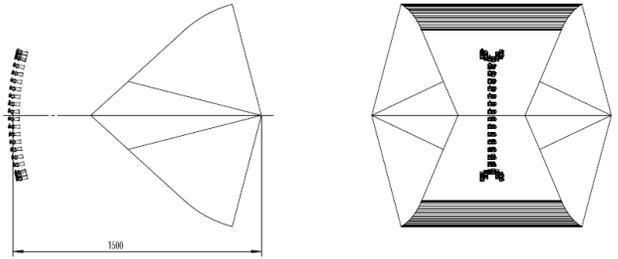

Fig. 2 The antenna design diagram.

The radar system together with the antenna array after assembly is pictured in Fig. 3.

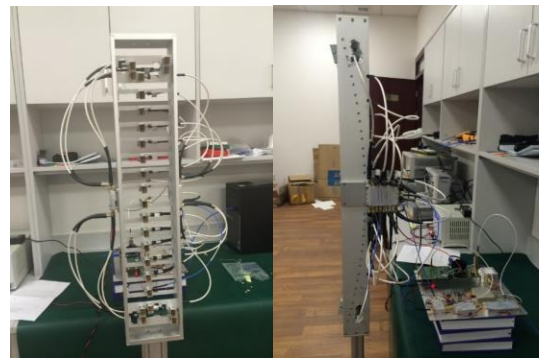

Fig. 3 The physical antenna array.

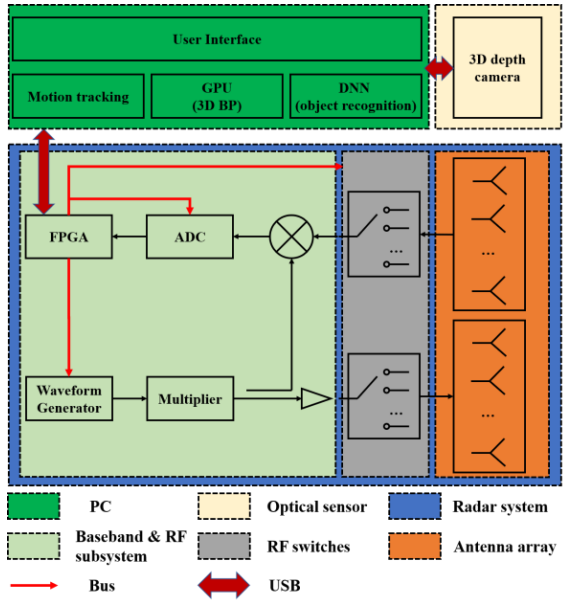

Fig. 1 System architecture.

## C. Baseband and RF Subsystems

The FMCW radar architecture is widely used due to its simplicity in system implementation and signal processing. The waveform generator is implemented using direct digital synthesis (DDS) technology. A 500MHz bandwidth chirp signal is generated from DDS and then up-converted to 3GHz before multiplied by 8 times to the K-band. Thus, the final transmitted FMCW signal ramps from 22GHz to 26GHz. The received signal is first amplified and then directly mixed with the reference signal which is coupled from the transmit channel to get the beat frequency signal at DC, which are then low-pass filtered and digitized.

The final output signal is measured to have output power 17.783dBm, bandwidth 22~26GHz, and the out-of-band suppression is about -45dB and the in-band ripple is within 3dB.

## D. System Timing

The system timing of a burst is illustrated in Fig .5. The radar system emits burst signals with pulse width of 30us and pulse recurrent time (PRT) of 40us. Switches are used to control the transmission and reception of signals to form 128 transmit-receive channels, which leads to a bust cycle of 5.12ms. The analog signals are converted to digital signals via the ADC and 201 points are sampled each pulse.

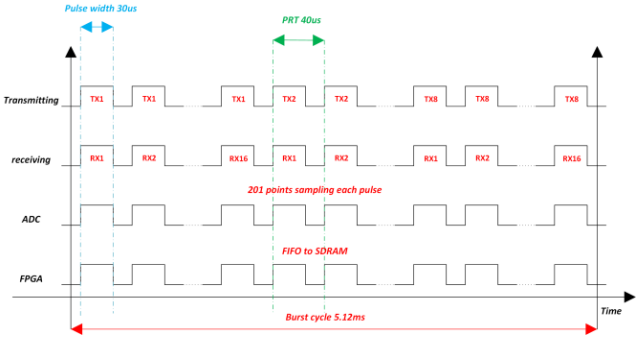

Fig. 5 The system timing of a burst.

## III. IMAGING ALGORITHM

### A. Array Calibration

The inconsistency across 16 Rx and 8 Tx channels will cause error in imaging and degrade the performance and quality of radar image. Such inconsistency is inevitable due to imperfect array installation, switch and cables, which will induce channel imbalances in terms of signal amplitude and phase as well as time delay. These errors would lead to image distortion and defocusing. The amplitude, phase imbalance and time delay differences across the channels must be calibrated for coherent radar imaging. The exact antenna phase centers should be estimated. TABLE II describes the parameters to be corrected for the MIMO array, the sources of errors, and their impact on imaging.

TABLE II The calibration parameters and effects on imaging

| parameters | error sources | image impact |
|---|---|---|
| channel amplitude imbalance | error from switch and cable | distortion |
| channel phase imbalance | error from switch and cable | defocusing |
| delay differential | error from switch and cable | defocusing |
| antenna phase center | array installation error | defocusing |

The array errors calibration methods can be classified into two categories: auto-calibration and active calibration [10]. We propose a MIMO array active calibration procedure based on a precision 2D linear stage. The 2D linear stage and the diagram of calibration setup is shown in Fig. 6. We use a long and phase-stable cable to connect the reference Tx (or Rx) port to an standard reference antenna which is mounted on the x-y 2D linear stage as shown in Fig. 6 (left). The linear stage used here has 0.02mm precision of positioning.

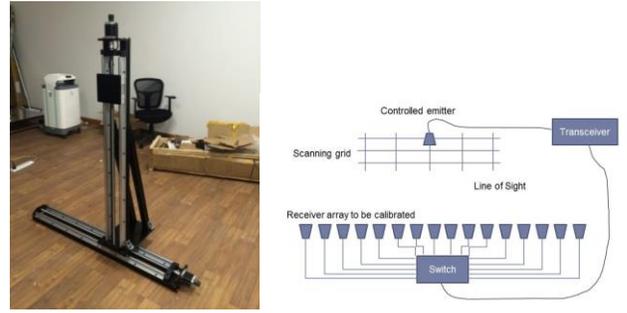

Fig. 6 (left) The physical view and (right) the measurement method of the scanning system.

During calibration, the 2D linear stage is controlled to scan across the grid of size 1mx1m with a step of 1cm. At each position, the received signal is recorded, which in fact measures the line-of-sight (LOS) link between the reference Tx (or Rx) antenna and one of the Rx (or Tx) antenna. When the LOS signal between the $i$ th antenna and the $j$ th point in the scanning grid is obtained and pulse compressed, the propagation delay $T_{ij}$, the peak amplitude $A_{ij}$ and the phase $\varphi_{ij}$ can be written as

$$T_{ij} = \frac{|\vec{p}_i - \vec{p}_j|}{c} + \tau_i \qquad (1)$$

$$A_{ij}\exp(j\varphi_{ij}) = \frac{a_i}{|\vec{p}_i - \vec{p}_j|^2}\exp\left[\frac{2\pi|\vec{p}_i - \vec{p}_j|}{\lambda} + \phi_i\right] \qquad (2)$$

where $a_i$, $\phi_i$ are amplitude imbalance and the phase imbalance of the $i$ th channel. $\tau_i$ is the time delay error and $\vec{p}_i$ is the phase center error of the $i$ th antenna. Using optimization techniques, the unknown parameters listed in TABLE II can be estimated. These parameters are then used to compensate the raw signals. As a demonstration, the imaging result of a point target (ball bearing) is shown in in Fig. 7. It plots the range vs array 2D real aperture imaging results before and after calibration. Apparently, the calibration has greatly improved the imaging results along the array dimension. After calibration, the imaging resolution is close to the theoretical value ~1.8 cm x 4 cm and the peak side lobe (PSL) is about -10 dB.

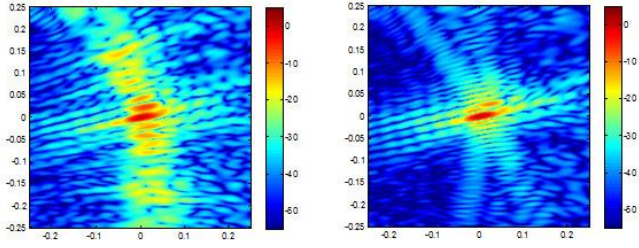

Fig. 7 The imaging result of a point target: (left) before calibration and (right) after calibration.

*B. Back projection on GPU*

Back projection (BP) algorithm is a typical time domain algorithm in radar imaging which coherently accumulates the echo from the same scatterer according to the back projection rule. As a time domain algorithm, BP has the advantage of flexibility to process any non-uniform array and synthetic aperture data. However, its disadvantage is the extremely high computational cost which renders it unacceptable in real-time applications such as security imaging. In this paper, we developed a 3D BP algorithm which can be executed in massive-parallel on GPUs.

3D BP algorithm uses the relative motion information of the platform and the target to calculate the distance history of scatterers located at different positions from the array. Then the complex phase history data is back projected to the imaging space, phase compensated and coherently accumulated. The return signal of the 2D planar MIMO array system can be represented by

$$s(x_t, y_t, x_r, y_r, k) = \iiint O(x, y, z) \exp(-jkR_t) \exp(-jkR_r) dx dy dz \quad (3)$$

where $R_t = \sqrt{(x_t - x)^2 + (y_t - y)^2 + (z_a - z)^2}$ represents the range from TX antenna to target area. $R_r = \sqrt{(x_r - x)^2 + (y_r - y)^2 + (z_a - z)^2}$ represents the range from RX antenna to target area. $z_a$ is the planar array position in $z$ axis. The TX and Rx antenna positions are represented as $(x_t, y_t, z_a)$ and $(x_r, y_r, z_a)$ respectively. $O(x, y, z)$ is the backscattering of target at position $(x, y, z)$. $k$ is the wavenumber. It can be solved using time domain integral method, written as

$$O(x, y, z) = \iiint s(x_t, y_t, x_r, y_r, k) \exp(jkR_t) \exp(jkR_r) dR_t dR_r dk \quad (4)$$

3D BP algorithm is a pixel-by-pixel imaging algorithm, where different pixels can be computed independently. This can be easily parallelized on GPUs. The rest computation such as FFTs can also be easily implemented. The CUBLAS and the CUFFT libraries are used to implement matrix operations and Fourier transforms respectively. Based on GPU and compute unified device architecture (CUDA) of NVIDIA corporation [11], 3D BP algorithm can be implemented. The flowchart of the GPU implementation is shown in Fig. 8. Compared with the processing time on CPU platform, the GPU implementation of 3D BP can be accelerated by over 400 times. With a powerful GPU station, it would take only 1 second to finish the imaging of one person.

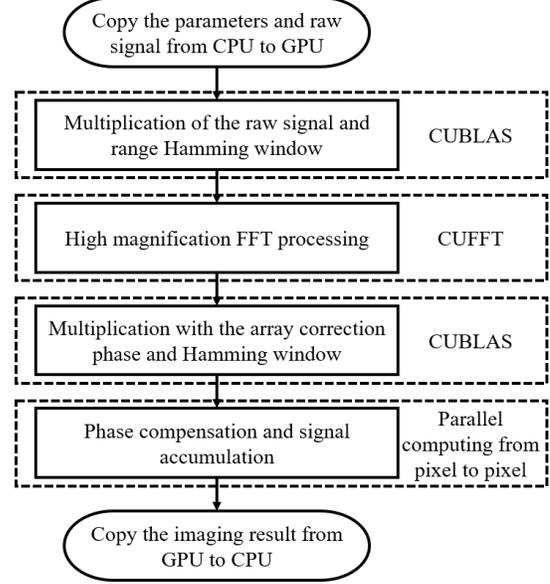

Fig. 8 The GPU implementation flow chart of 3D BP algorithm.

*C. Auxiliary sensor for motion tracking*

In ISAR imaging, the target motion parameters cannot be known preliminarily and estimated from the echo signals [12][13][14]. However, these methods only work with the target of dominated scatterers which are not suitable for the human body ISAR imaging. Then, we decompose the human motion model with 20 key components and conduct optical sensor based tracking and reconstruction of the human motion.

The tracking optical device used in this paper is Kinect. The raw data error of human body positioning is measured to be around 1cm.

Then Kalman filter is used to filter jumping faults of the measurements. When the measured value is greatly different from the predicted value of Kalman filter, it is filtered. After the MRA fitting and Kalman filtering, the residual errors in $x$ direction, $y$ direction and $z$ direction are 1.15 mm, 1.17 mm and 2.26 mm respectively.

## IV. EXPERIMENT RESULTS

*A. Human Imaging*

To verify the imaging performance of the security system, the experiments of human body carrying different objects are carried out with the real system. Since the motion errors cannot be completely eliminated, motion information is measured with laser rangefinder or Kinect. Human body stands on a rail car moving across the surveillance area to be imaged and checked. The experiment scene is shown in Fig. 11.

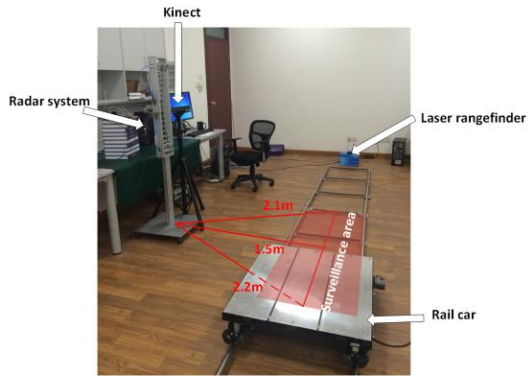

Fig. 11 The experiment scene of human imaging.

The human body imaging results of laser rangefinder and Kinect are shown in Fig. 12. It is depicted that the system can effectively work with an auxiliary sensor like rangefinder and Kinect. However, Kinect can provide human body motion information of more than 20 joints directly while rangefinder cannot.

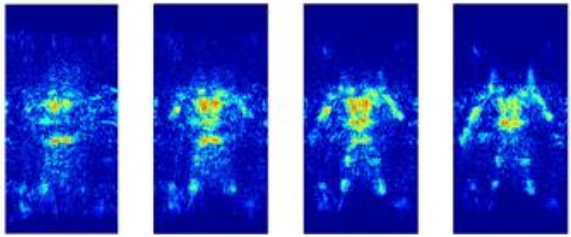

Fig. 12 The human body imaging results of lase rangefinder and Kinect.

Using Kinect, the human body carrying nothing, cellphone and gun are imaged as shown in Fig. 13. The images are generated using 3D BP on GPU and the image slices of different distances are tiled and listed.

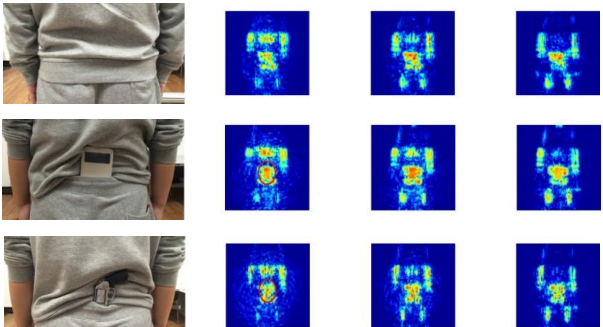

Fig. 13 The imaging results of human body carrying different objects.

### B. Object Recognition

Studied from some researches, the automatic target recognition (ATR) methods can be used in radar target classification [15][16][17][18]. Moreover, the ATR method can be also used in security screening and provide a function to automatically determine whether an anomaly is present to effectively avoid the problem of "invasion of public privacy".

ATR with computer is one of the key technologies to realize the object recognition in this paper. We use MIMO antenna to obtain the scattering information of the observed target. Derived from the high resolution millimeter wave images, the size, shape and EM characteristics of the target can be obtained. ATR can be better implemented with these features. We use deep neural networks (DNN) for ATR and the design of DNN diagram is shown in Fig. 14.

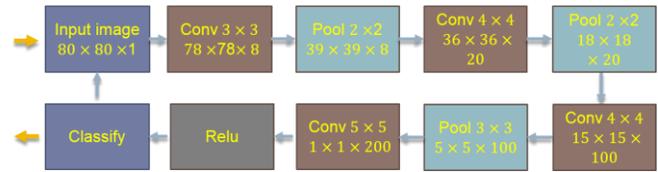

Fig. 13 The designed DNN diagram.

We use the millimeter wave images of the proposed system as training set. The training set is extended by means of translation, rotation and noise addition of the images. Finally, the object recognition experiments are taken out. The human body with and without items are imaged and recognized. When the human body passes through the inspection area without concealed targets, repeated tests are carried out to calculate false alarm rate (FAR). The FAR is about 8% when the tests are repeated 20 times.

The missing alarm rate (MAR) and the recognition accuracy are also calculated by repeated tests. The human body conceals and carries three different objects close to the body and passes through the inspection area for repeated testing. After each situation is repeated 20 times, the calculated MAR is 0.3% and the recognition accuracy is about 96%.

## V. CONCLUSION

Aimed at the human body microwave and millimeter wave security imaging, a 3D ISAR system at 24GHz with a sparse MIMO array is proposed in this paper. The system design and imaging algorithm with this system is described. The human body is imaged using a quasi-real-time 3D BP algorithm with acceleration of GPU. Radar images of the human body are automatically detected with DNN whether prohibited objects are concealed. A novel array calibration approach is developed to calibrate the channel imbalance across the MIMO array. Finally, experiment results of human imaging and object recognition are provided to prove the efficiency of the proposed security system. The auxiliary target tracking information is used to combine MIMO and ISAR technologies, bringing out an effective way to realize the security imaging, while the ATR method with DNN effectively avoids the privacy invasion. Therefore, this system is reasonable and suitable to be used in the security field.

## REFERENCES

[1] J. Nation, W. Jiang, "The utility of a handheld metal detector in detection and localization of pediatric metallic foreign body ingestion," International journal of pediatric otorhinolaryngology, vol. 92, pp. 1-6, 2017.

[2] J. Gao, Y. Qin, B. Deng, et al., "A novel method for 3-D millimeter-wave holographic reconstruction based on frequency interferometry techniques," IEEE Transactions on Microwave Theory and Techniques, vol. 66, no. 3, pp: 1579-1596, 2017.

[3] J. Gao, B. Deng, Y. Qin, et al., "An Efficient Algorithm for MIMO Cylindrical Millimeter-Wave Holographic 3-D Imaging," IEEE Transactions on Microwave Theory and Techniques, no. 99, pp: 1-10, 2018.

[4] S. Degirmenci, D. G. Politte, C. Bosch, et al., "Acceleration of iterative image reconstruction for x-ray imaging for security applications," Computational Imaging XIII. International Society for Optics and Photonics, vol. 9401, pp: 94010C, 2015.

[5] I. Uroukov, R. Speller, "A preliminary approach to intelligent x-ray imaging for baggage inspection at airports," Signal Processing Research, vol. 4, pp: 1-11, 2015.

[6] L3 Security & Detection Systems Inc., "provision2 factsheet," L3 SDS, Woburn, MA, USA. [Online]. Available: https://storage.pardot.com/16582/113781/PROVISION2_FACTSHEET_23MAR17_PF.pdf.

[7] A. Schiessl, A. Genghammer, S. S. Ahmed, and L. P. Schmidt, "Hardware realization of a 2 m $\times$ 1 m fully electronic real-time mm-wave imaging system," in Proc. European Conference on Synthetic Aperture Radar, pp. 40-43, 2012.

[8] S. S. Ahmed, A. Schiessl, F. Gumbmann, and M. Tiebout, "Advanced Microwave Imaging," IEEE Microwave Magazine, vol. 13, no. 6, pp. 26-43, 2012.

[9] S. S. Ahmed, "Electronic microwave imaging with planar multistatic arrays," Logos Verlag Berlin GmbH, 2014.

[10] D. Wang, Y. Wu, "Array errors active calibration algorithm based on instrumental sensors," Science China Information Sciences, vol. 54, no. 7, pp: 1500-1511, 2011.

[11] T. Soyata, "GPU Parallel Program Development Using CUDA," Chapman and Hall/CRC, 2018.

[12] C. M. Ye, J. Xu, Y. N. Peng, et al., "Key parameter estimation for radar rotating object imaging with multi-aspect observations," Science China Information Sciences, vol 53, no. 8, pp: 1641-1652, 2010.

[13] L. Zhang, Y. C. Li, Y. Liu, et al., "Time-frequency characteristics based motion estimation and imaging for high speed spinning targets via narrowband waveforms," Science China Information Sciences, vol. 53, no. 8, pp: 1628-1640, 2010.

[14] S. Shao, L. Zhang, H. Liu, et al., "Spatial-variant contrast maximization autofocus algorithm for ISAR imaging of maneuvering targets," Science China Information Sciences, vol. 62, no. 4, pp: 40303, 2019.

[15] K. El-Darymli, E. W. Gill, P. Mcguire, et al., "Automatic target recognition in synthetic aperture radar imagery: A state-of-the-art review," IEEE Access, vol. 4, pp: 6014-6058, 2016.

[16] S. Chen, H. Wang, F. Xu, et al., "Target classification using the deep convolutional networks for SAR images," IEEE Transactions on Geoscience and Remote Sensing, vol. 54, no. 8, pp: 4806-4817, 2016.

[17] Z. Huang, Z. Pan, B. Lei, "Transfer learning with deep convolutional neural network for SAR target classification with limited labeled data," Remote Sensing, vol. 9, no. 9, pp: 907, 2017.

[18] J. Zhao, W. Guo, Z. Zhang, et al., "A coupled convolutional neural network for small and densely clustered ship detection in SAR images," Science China Information Sciences, vol. 62, no. 4, pp: 42301, 2019.